\documentclass{aa}
\usepackage{graphicx,float}
\def\msun{~M$_\odot$}
\def\kms{~km~s$^{-1}$}
\begin{document}

\title{The ionising cluster of 30 Doradus
\thanks{Based on observations collected at the European Southern Observatory}}
\subtitle{IV. Stellar kinematics}
\author{
  G. Bosch\inst{1,2}\and
  F. Selman\inst{3}\and
  J. Melnick\inst{3}\and
  R. Terlevich\inst{1}\thanks{Visiting Professor, INAOE. M\'exico}
  }
        
\offprints{guille@fcaglp.unlp.edu.ar}

\institute{
  Institute of Astronomy, Madingley Road, Cambridge CB3 0HA \and
  Facultad de Ciencias Astron\'omicas y Geof\'{\i}sica, La Plata, Argentina \and
  European Southern Observatory, Alonso de C\'ordova 3107, Santiago, Chile
  }

\date{Received 2001; accepted 2001}

\abstract{
On the basis of 
multislit spectroscopy of 180 stars in the ionising cluster of 30 
Doradus we present reliable radial velocities for 55 stars.
We calculate a radial velocity dispersion of $\mathbf{\sim35}$\kms\  
for the cluster and we  analyse the possible influence of spectroscopic 
binaries in this rather large velocity dispersion. We use numerical simulations to
show that the observations are consistent with the hypothesis that all the stars
in the cluster are binaries, and the total mass of the cluster is
$\sim 5 \times 10^{5}$\msun.  A simple test shows only marginal evidence for
dynamical mass segregation which if present is most likely not due to dynamical
relaxation.
}

\maketitle

\section{Introduction}

A significant fraction of the old stars we now observe in galaxies
appear to have formed in Starbursts. Therefore, understanding violent
star formation becomes crucial if we want to understand
the star formation history of the Universe. 30~Doradus in the LMC is
the nearest and best studied example of a massive starburst cluster and
thus it has become a sort of `Rosetta Stone' for deciphering the physics of
violent star formation processes (Walborn \cite{walborn91}, Selman et al. 
\cite{1999A&A...347..532S} (Paper III) and references therein).

Although 30~Dor has been the subject of intensive observational effort
from the ground and space 
there are still a number of critical problems that remain unsolved.
Perhaps the most burning open problem is the issue of mass segregation 
first raised by Malumuth \& Heap (\cite{malumuth95}) that
has  important implications
for our understanding of the process of cluster formation in general
(Clarke \cite{clarke01}).

This Paper is the fourth of a series devoted to a comprehensive
study of the 30~Doradus starburst cluster (NGC~2070). 
We can summarize the central results of the previous papers of this
series as follows:

\begin{itemize}
	\item The Initial Mass Function (IMF) of NGC~2070 between 3 and 120\msun\ 
	is very well represented by a Salpeter power-law.  The total
	`photometric' mass of the cluster -- inside 20 pc -- from 0.1\msun\ 
	to 120\msun\ is $M_{\rm{phot}}\sim 3 \times 10^5$\msun.

	\item The star-formation history of the cluster shows three distinct
	peaks centered 5~Myrs, 2.5~Myrs, and $<1.5$Myrs ago with no evidence for a spatial
	segregation of ages.

	\item The density distribution of stars between 20 and 60\msun\ is
	a power-law of slope of $-2.85$. This means that the total mass increases 
	very slowly with radius ($r^{0.15}$ or even logarithmically.

	\item There is some evidence for mass segregation revealed by the
	fact that the most massive stars appear to be more concentrated toward
	the center. However, there is a distinct `ring' of massive stars located
	at about 6pc from the center of the cluster (R136). This ring contains
	stars of the three age groups described above.
\end{itemize}
The purpose of the present Paper is to study the stellar kinematics
of the brightest stars in the 
cluster using the NTT observations described in Paper~II of this series 
(Bosch et al. \cite{1999A&AS..137...21B}). We use the data to 
investigate the presence of dynamical mass
segregation in the cluster, and to attempt to constrain the lower end of the
IMF from the comparison between the photometric and the dynamical masses.

\section{Observations and Data Reduction}
\label{sec:rvobs}

The observations -- already described in Paper II --
were obtained with the La Silla NTT telescope using the Multislit option of EMMI's 
RILD mode. In this mode, grism \# 5 was used, which in combination with
a $0.8$ arcsec wide slit yields a resolution of 1.3 \AA~/~pixel. The 
wavelength range is 3600 to 6000 \AA~in most cases,
although the limits vary slightly between individual spectra, as they
depend on the position of each slitlet within the mask.
A total of 7 masks were 
produced each including an average of 35 
slitlets. 
The basic reduction steps were described in Paper~II including bias
correction, flat fielding, wavelength calibration, and aperture extraction.
Here we need to discuss some aspects of the wavelength calibration not included in Paper~II
that are particularly relevant for the determination of radial velocities.


During the observations a technical problem made the 
Argon lamp too weak to produce reliable identification lines 
{\bf for 2 masks} while the strong \ion{He}{i}~5875 \AA~ {\bf arc} line is 
saturated for several spectra. The problem is worst when both
effects appear at the same time as there are no reliable comparison lines
redwards \ion{He}{i}~5015 \AA. For example, this makes useless the 
\ion{He}{ii}~5411 \AA~absorption line for radial velocity measurements in the
affected spectra.

The problems with the calibration lamps also altered the
normal sequence in which the lamps were taken; the calibration exposures
{\bf for the last three masks} had to be taken at the end of the night.
Although there were no telescope presets, the telescope was still tracking
so the position angle of the instrument rotator slowly changed and this
may introduce systematic errors.  We address this issue in the following
section.

\section{Radial Velocities}
\label{sec:rvcal}

\subsection{Zero-point errors}
\label{subsec:zeropoint}

The problems with the calibration lamps may introduce systematic
differences between the zero points of each mask. We have used 
the radial velocity of the  [\ion{O}{i}] 5577 \AA\  Auroral line, 
whenever possible, to test for this effect. 

The straightforward approach would be to add algebraically 
the radial velocity of the sky line to
the velocity of each star. {\bf Unfortunately, however,
the problems with the calibration lamps described above imply that
several spectra have unreliable wavelength
calibration in the region of [\ion{O}{i}].}

For most spectra the formal (measurement) error in the position of
the sky line is $\sim$6,6\kms.  The
mean radial velocity of the [\ion{O}{i}] lines with reliable calibrations is
-6.5\kms\ with a dispersion 
of 12.9\kms.  Thus,
while we cannot correct the individual velocities for systematic zero point
errors, the radial velocity dispersion must be corrected by this effect by
subtracting quadratically the [\ion{O}{i}] dispersion $\sigma_{[\ion{O}{i}]}=\sqrt{12.9^2 - 6.6^2}=10.9$\kms.

\subsection{Velocity determination}

As mentioned above, the resolution of our combination of dispersion
grating and camera yields 1.3~\AA~per pixel. For  2-pixel sampling 
this corresponds to a spectral resolution of about 165\kms\ at 4750\AA.  
The ultimate limit
attainable in the precision of  Doppler shifts is dominated by
the photon noise in the spectrum 
(Brown \cite{1990ccda.proc..335B}). 
The uncertainty in the
measured radial velocity for the case of a
single line of width $w$ and depth $d$, measured in units of the
continuum intensity $I_{\rm{c}}$, is given by
\begin{equation}
\delta v_{\rm{rms}} = \frac{c \, w}{\lambda \, d \, (N_w \, I_{\rm{c}})^{1/2}}
\end{equation}
where $N_w$ is the number of samples across
the width of the line, 
$c$ is the speed of light, and $\lambda$ is the central wavelength
of the line. For typical lines measured
in our stars ($\lambda = 4500$\AA, $w$=7\AA, $N_w$=6, $d$=0.2, and
$I_{\rm{c}} = 2\times10^4$), we obtain $\delta v_{\rm{rms}}$ = 15 km s$^{-1}$.
This is {\bf smaller} than the velocity dispersion expected if the cluster
is virialised and the total mass is close to our photometric estimate.
However, the signatures of expanding stellar atmospheres 
and binaries may be much stronger than the virial motions. Thus, it is
very  important
to constrain these effects with the data at hand 
before embarking in a high spectral resolution
survey of the kinematics of the cluster.

At our resolution the Balmer lines cannot be used to measure radial velocities
because they are severely contaminated by nebular emission.  Therefore, we have
restricted our analysis to stars with well exposed HeI and HeII absorption 
lines.
In order to have an indication of the presence of atmospheric motions we
have only considered stars with at least three He lines detected.  This 
further restricts
our sample to 97 spectra, several of which correspond to the same star.

The centroids of the lines were determined from Gaussian fits using the package
\tt ngauss \rm within IRAF.  The fitting errors were used as weights to
calculate the (weighted) mean velocity of each star.  A conservative 
$\kappa-\sigma$ filter was used to reject stars with suspected internal (atmospheric)
motions. Thus, all stars with a dispersion of more than 25\kms\ between the
measured lines were rejected.  The final list is presented
in Table~\ref{final} that gives, for each star, the Parker number, the spectral
type from Paper~II, position in arc-sec
from the cluster center, assumed to be R136 (Selman et al. 
\cite{1999A&A...341...98S}), the weighted average
radial velocity, and the weighted error. 
{\bf A number of stars appear to be
binaries on the basis of showing double peaked lines (stars \# 1024, 1369 and 
1938), asymmetric line profiles (\# 222, 613, and 1191), 
or different radial velocities for the \ion{He}{i} 
and \ion{He}{ii} lines (\# 1998).}
These stars tend to have larger internal errors as can be seen in the second
part of
Table~\ref{final}.

\begin{table}[h!]
\caption{Stellar radial velocities}
\label{final}
\begin{center}
\begin{tabular}{lrrrrr}
\hline\hline
Star id.\ & Sp.Type & $\Delta\alpha('')$ & $\Delta\delta('')$ &~~$\langle V_{r} \rangle$ & $\sigma_{\rm{int}}$   \\
\hline
    15 & O8.5~V        &-107 &   107 & 234.9        & 14.0 \\
    32 & O9~IV         &-102 &    72 & 272.1        & 09.5 \\    
   124 & O8.5~V        & -76 &    25 & 287.5        & 06.9 \\    
   316 & O6.5 V        & -50 &  -164 & 265.9        & 09.5 \\   
   541 & O7.5 V        & -29 &   -65 & 252.6        & 07.1 \\   
   649 & O8-9 V        & -20 &  -106 & 323.7        & 09.4 \\    
   713 & O5 V          & -15 &   -53 & 308.7        & 11.8 \\   
   747 & O6-8 V        & -13 &  -142 & 364.3        & 22.8 \\    
   791 & O5 V          & -09 &   141 & 310.7        & 08.8 \\    
   805 & O5-6 V        & -08 &   -38 & 292.1        & 09.4 \\    
   863 & O6.5 V        & -04 &   -03 & 308.0        & 06.5 \\    
   871 & O4 V((f$^*$)) & -04 &   -44 & 290.3        & 06.5 \\   
   901 & O3 V((f$^*$)) & -02 &    26 & 276.2        & 08.7 \\    
   905 & O9-B0 V       & -02 &    61 & 198.2        & 16.3 \\   
   975 & O6-7 V((f))   &  02 &   -27 & 325.5        & 05.9 \\    
  1022 & O5: V         &  04 &  -139 & 320.6        & 07.2 \\   
  1063 & O6-7 V        &  06 &   108 & 267.0        & 16.5 \\    
  1109 & O9 V          &  09 &  -167 & 229.7        & 05.9 \\   
  1139 & B0 V          &  11 &    36 & 225.5        & 09.5 \\   
  1163 & O4 If:        &  12 &   -72 & 274.1        & 11.3 \\   
  1247 & B0.5 IV       &  17 &    91 & 333.0        & 11.6 \\   
  1283 & O6 V:((f$^*$))&  19 &   -09 & 352.2        & 06.7 \\    
  1339 & B0-0.2 IV     &  23 &   -60 & 265.5        & 12.0 \\    
  1389 & B1: V::       &  27 &    70 & 292.6        & 06.1 \\       
  1419 & B0-0.2 III-I  &  31 &    98 & 259.0        & 10.0 \\        
  1459 & O9.5 II       &  34 &   145 & 272.7        & 14.3 \\        
  1460 & B0-2 V        &  34 &   172 & 282.7        & 21.0 \\        
  1468 & O9.5 V        &  36 &    16 & 321.0        & 13.3 \\        
  1500 & B0.2 III      &  39 &    40 & 275.2        & 09.0 \\        
  1531 & O6 V((f))     &  43 &   -25 & 308.0        & 09.0 \\       
  1553 & O7 V          &  47 &   -09 & 321.5        & 07.3 \\        
  1563 & O7.5 II-III(f)&  47 &   -04 & 271.6        & 06.9 \\        
  1584 & B0-1 V        &  50 &   -02 & 320.8        & 23.1 \\        
  1604 & B1 V          &  55 &    85 & 360.2        & 17.6 \\       
  1614 & O5-6 V((f))   &  56 &    09 & 291.2        & 06.6 \\        
  1618 & B0-0.2 III    &  56 &   128 & 270.1        & 07.6 \\       
  1619 & O8 III(f)     &  56 &   102 & 357.8        & 20.0 \\        
  1643 & O5 V          &  60 &   128 & 279.0        & 06.6 \\        
  1661 & B1 III        &  62 &   124 & 322.7        & 07.5 \\        
  1685 & B0.5-0.7 III-I&  66 &   161 & 291.8        & 11.6 \\       
  1729 & B1 II-III     &  71 &    80 & 283.2        & 17.3 \\      
  1737 & B1.5 III      &  71 &   139 & 339.1        & 05.7 \\       
  1969 & B0.7 IV       & 113 &    74 & 329.6        & 12.7 \\        
  1987 & B2 I          & 120 &  -113 & 294.8        & 05.0 \\        
 10001 & O4 V          &     &       & 246.1        & 14.7 \\       
 10003 & B1-1.5 V      &     &       & 279.9        & 06.9 \\
\hline
       \multicolumn{6}{c} {Suspected Binaries} \\ \hline
  222  & O9.5-B0 V     & -62 &   143 & 198.9 & 26.4 \\
  613  & O8.5 V        & -23 &  -154 & 203.2 & 13.7 \\
 1024  & O9-B0 V       &  05 &  -110 & 510.7 & 27.9 \\
 1191  & B0.2-1 III-V  &  13 &   -30 & 345.7 & 28.8 \\
 1369  & O8.5 V        &  26 &   -09 & 318.1 & 78.5 \\
 1938  & O7.5 V        & 107 &   134 & 350.0 & 18.0 \\
 1988  & B0.5 V        & 121 &   -22 & 300.0 & 18.0 \\ 
\hline\hline
\end{tabular}
\end{center}
\end{table}

\begin{figure}[h!]
\begin{center}
\resizebox{.95\hsize}{!}{\includegraphics{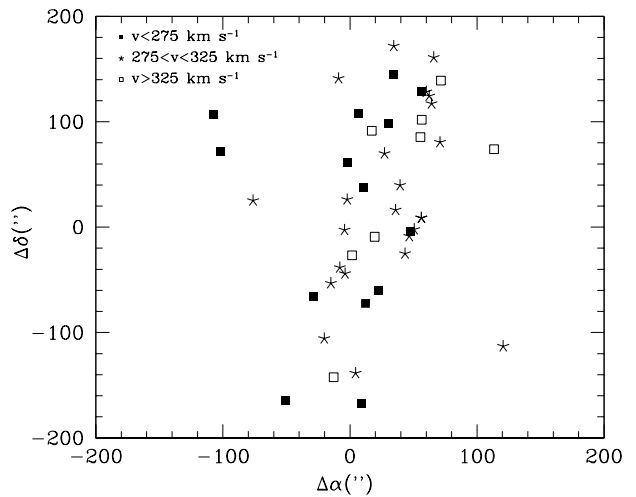}}
\caption{Spatial distribution of stellar radial velocities. Two stars
are missing, as they fall out of the photometry area}
\label{spacedist}
\end{center}
\end{figure}

\section{Results}
\subsection{The distribution of radial velocities}

In Figure \ref{spacedist} we plot the spatial distribution of the measured 
stars in three different velocity bins in order to check for the presence
large scale motions, such as rotation, or clustering of stars
in large substructures of different kinematics. 
Unfortunately, the distribution of the points, dictated by the geometry of
the spectral masks, precludes a finer analysis, but a visual inspection of the
graph reveals no evidence of clumping of stars with similar velocities, nor of rotation
along the axis defined by the observational technique.

\begin{figure*}
\begin{minipage}[t]{.45\textwidth}
\includegraphics[width=8.5cm]{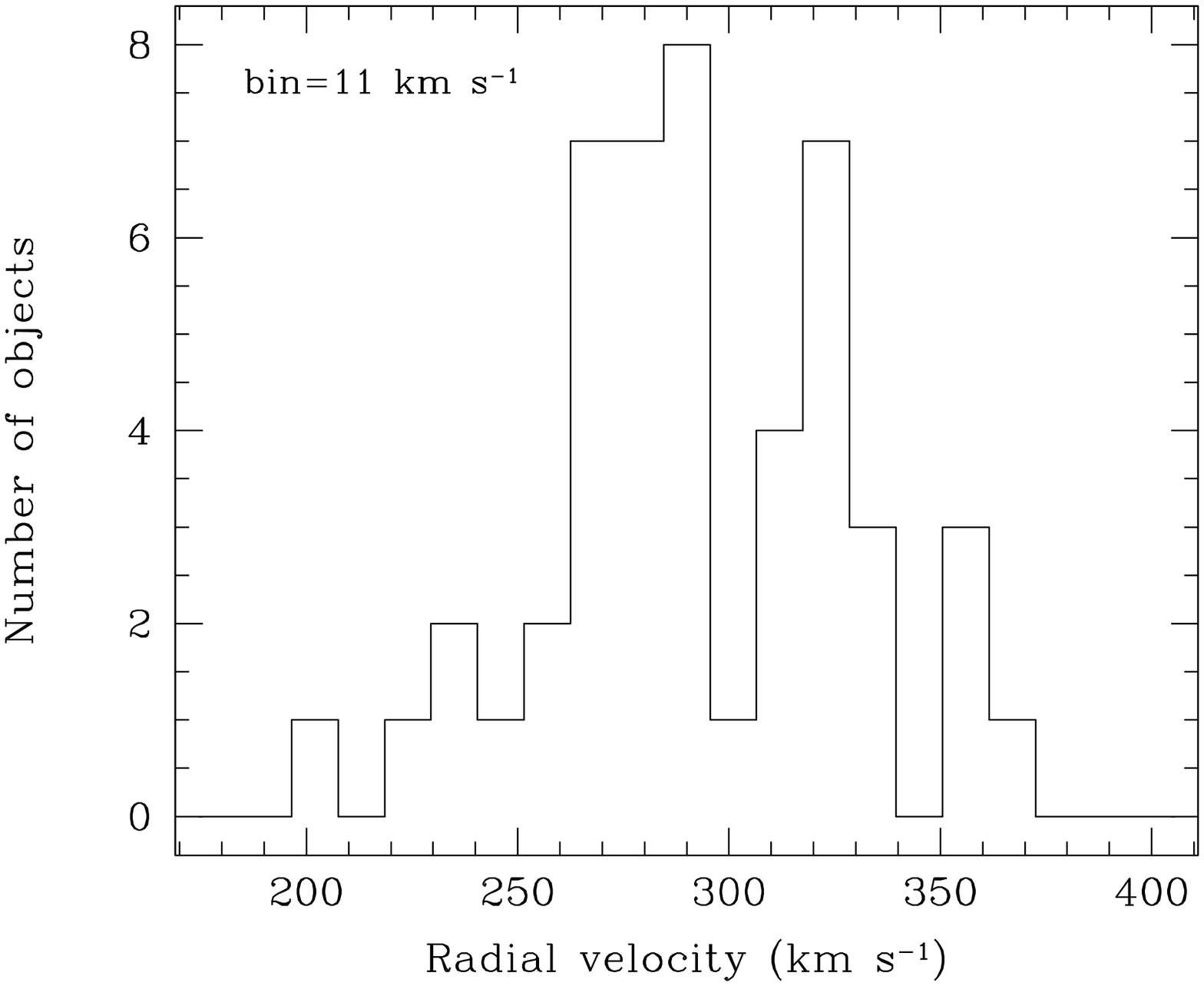}
\end{minipage}
\hfill
\begin{minipage}[t]{.45\textwidth}
\includegraphics[width=8.5cm]{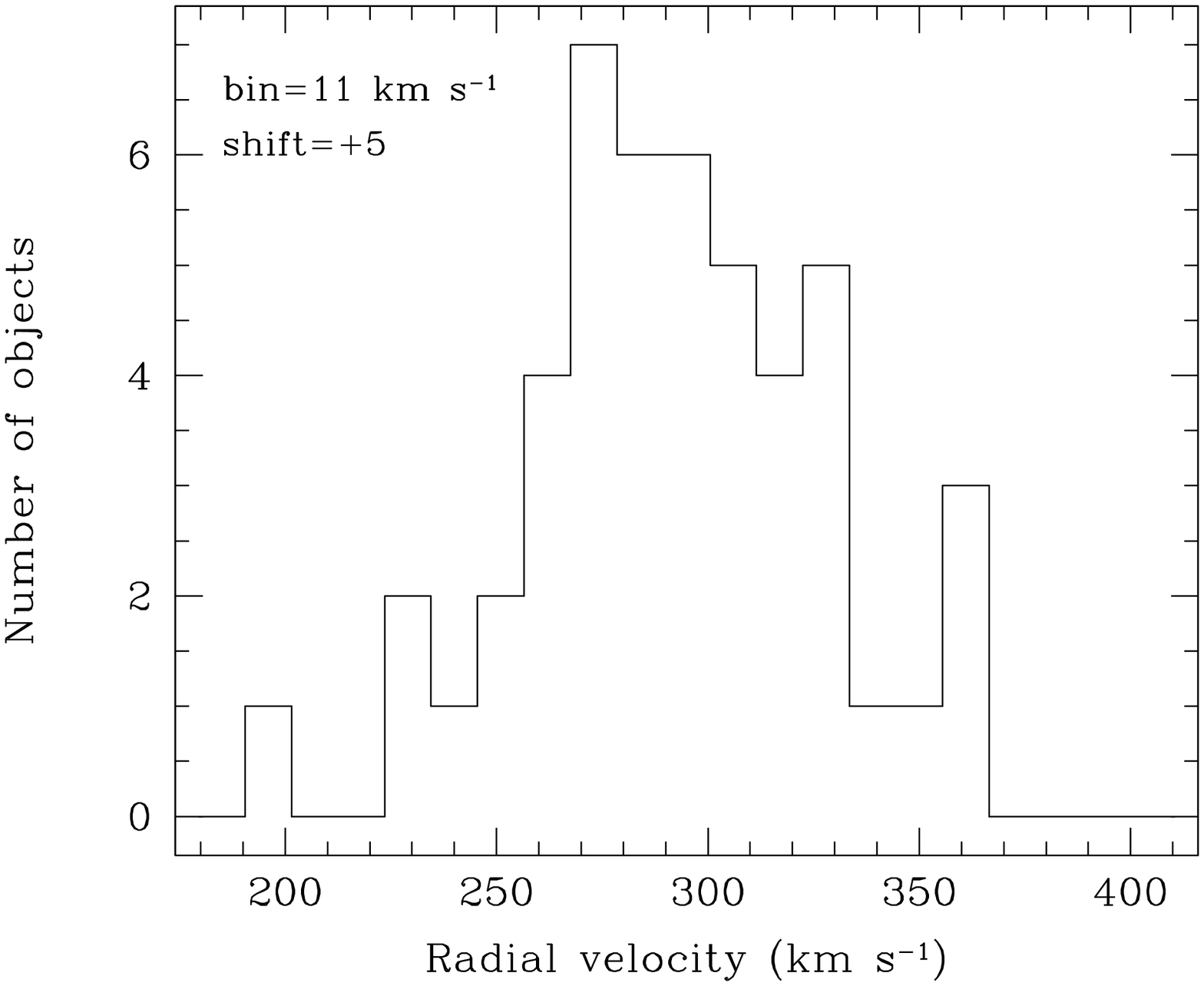}
\end{minipage}\\
\begin{minipage}[t]{.45\textwidth}
\includegraphics[width=8.5cm]{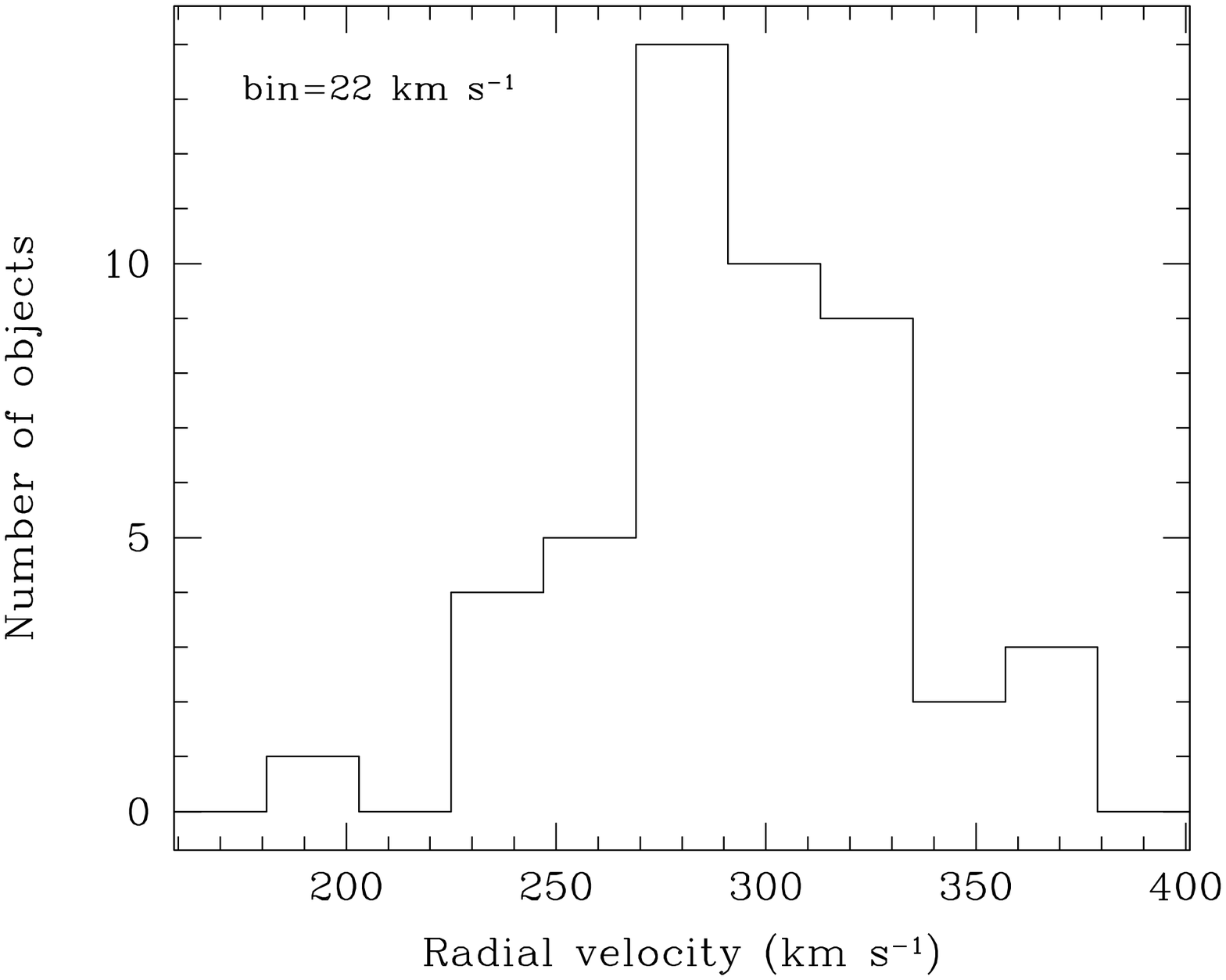}
\end{minipage}
\hfill
\begin{minipage}[t]{.45\textwidth}
\includegraphics[width=8.5cm]{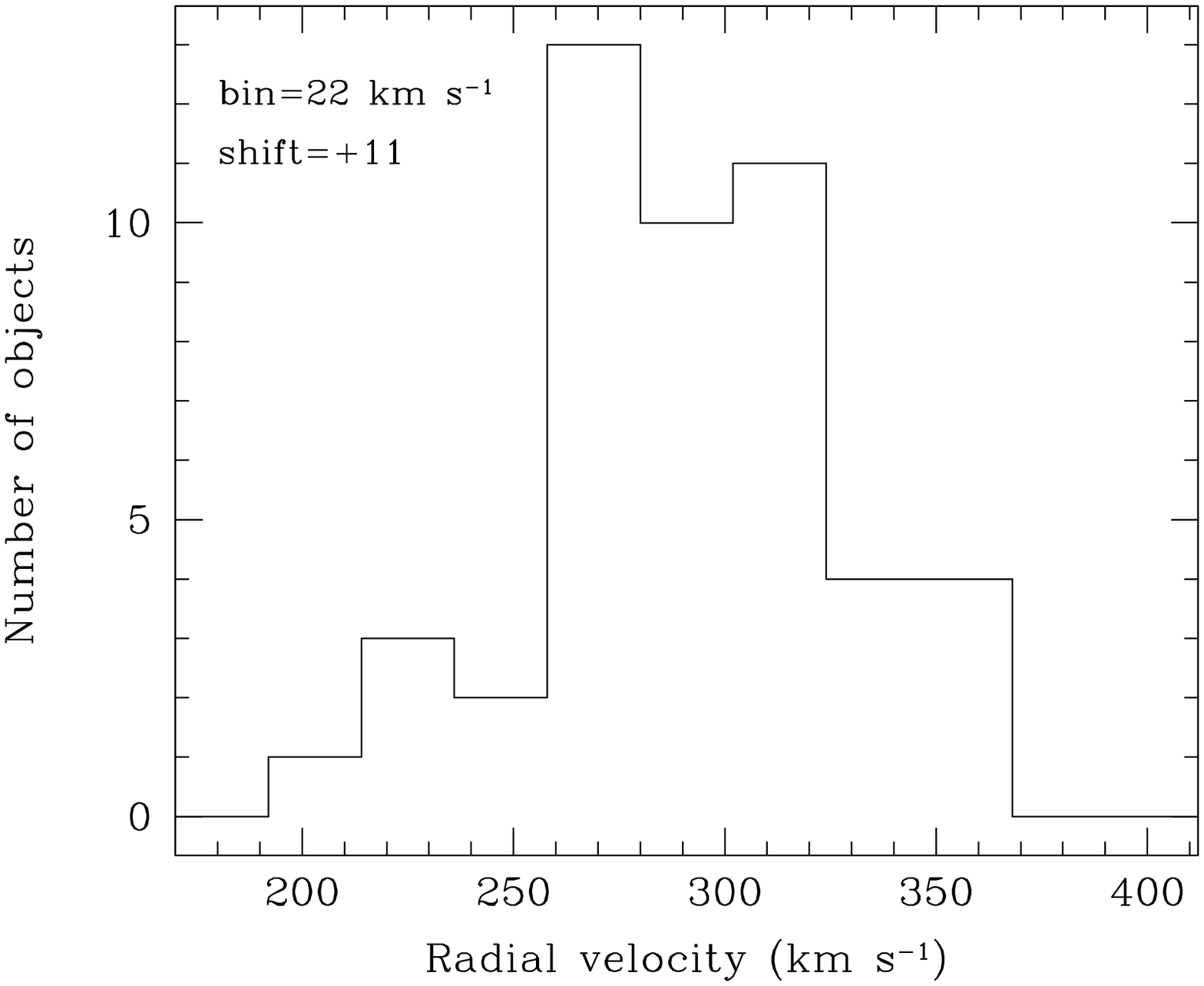}
\end{minipage}
\caption{Distribution of radial velocities for two bin sizes of corresponding to
the typical measurement error (11\kms, top panels) and twice that error (bottom panels).
The histograms are shifted by half a bin between the left and the right panels to
illustrate features due to sampling statistics}
\label{hist1}
\end{figure*}

We can safely proceed, therefore, to draw histograms and estimate
velocity dispersions.  Figure~\ref{hist1} presents velocity histograms of
the single stars in Table~\ref{final} with two different bin sizes, and two
different origins to check for sampling effect.
The top two panels are for a binning of 11\kms\ (corresponding
to our estimate for the mean random error in the velocities) and the lower two for
a binning of twice this error.  The multiple peaks of the first plot disappear when
the bins are shifted by 5\kms\ (half a step), indicating that they are artifacts of the
small number statistics.  This is confirmed in the lower panel where
shifting the sampling by half a step (11\kms) does not change the distribution
in any significant way. The hypothesis of a Gaussian distribution is valid,
based on $\chi^2$ tests performed to the distribution.

We conclude
that there is no  evidence for statistically significant 
peaks in the radial velocity distribution of the cluster. Thus, we can use all the data
to estimate the velocity dispersion of the cluster.
After correction for measurement (internal)
errors and zero point errors (from the [\ion{O}{i}] Auroral line), the 
radial velocity dispersion of 48 single stars in the cluster is 32\kms.
This is much larger that the value expected if the cluster is
virialised with a total mass equal to the photometric mass and also much larger
than our combined errors ($\sigma_{\rm{tot}}\sim 15$\kms).

\subsection{Binaries}

Spectroscopic binaries are very difficult to detect in a single observation,
specially single-lined ones. Clearly, therefore, the number {\bf of} binary 
candidates 
listed in Table~\ref{final} is only a \it lower limit.\rm This is consistent 
with the fact that studies of young open clusters indicate 30\% to be a 
typical percentage of binaries detected in these systems (Garmany et al.\
\cite{1980ApJ...242.1063G}, Levato et al. \cite{1991Ap&SS.183..147L}) 
to be compared to 13\% in our sample.  

Our data alone, therefore, can only be used to determine a lower-limit 
to the effect 
of binaries in the observed velocity dispersion of NGC~2070.  
An upper limit can be obtained using the Montecarlo simulations of 
Bosch and Meza (\cite{boschmeza98}). Assuming that all the stars in the 
cluster are binaries with their center of mass 
at rest within the cluster 
the models predict a velocity dispersion of $\sigma_{\rm{bin}} \sim 35$\kms.
This must be compared with our observed dispersion of 46.5\kms including the 55 stars
of Table~\ref{final} (corrected for observational errors).
If we exclude star \#1024 with a radial velocity of 510\kms
which may not be a member of the cluster, the dispersion is reduced
to 36.5\kms.  This result is  
 consistent, within the uncertainties,
with the hypothesis that most of our observed velocity dispersion for the cluster is
due to binary motions.  A (very uncertain) {\bf lower} limit for the virial motions
of the stars in the cluster potential is thus,
$\sigma_{\rm{vir}}=\sqrt{36.5^2 - 35^2}\sim 10$\kms.

{\bf For comparison purposes we can estimate the expected velocity dispersion
assuming the cluster is virialised and the total mass is equal to the upper
photometric mass limit from Paper III. From the density distribution
derived in the same paper, we estimate a core radius of 0.5 pc which yields}
$\sigma_{\rm{phot}}=18$\kms.

\subsection{Mass Segregation}

In Paper II, using only the stars with spectroscopy, we found that the most
massive stars in NGC~2070 were preferentially found closer to the center
of the cluster. This was interpreted as tentative evidence in favor of the
existence of mass segregation, as was originally advocated
by Malumuth and Heap (\cite{malumuth95}). This conclusion was somewhat 
weakened 
in Paper~III which presented a detailed analysis of the IMF in several rings
around the cluster center. The IMF was found to have the Salpeter slope 
almost everywhere with the exception of the very core where, 
combining intermediate mass HST data from Hunter et al. (\cite{hunter95}) 
with our high mass end data, we found marginal evidence for flattening. 
The most important "mass segregation" was found in a "ring" 6 pc away from 
the cluster center, again weakening the idea that closer to the center
we would find the major relative concentration of massive stars. We should point
out in the context that,
because of the strong density gradient, the half-mass ratio of the cluster
is very small. This explains the large concentration of high mass stars in the 
central parsec of the cluster found by Massey and Hunter (\cite{massey98}).

The two-body relaxation time for NGC~2070 
\footnote{Although for systems with a 
wide spectrum of masses the relaxation times are much 
shorter (Inagaky and Saslaw \cite{1985ApJ...292..339I}).}
is about two orders of magnitude
larger than the age of the stars. Therefore, if mass segregation is indeed
present it must be primordial (Bonnel \& Davies \cite{1998MNRAS.295..691B}, 
Elmegreen \cite{elmegreen00}).
In either case, dynamical or primordial, we expect to see a difference in the
velocity dispersion of the stars as a function of mass in the sense of it being
lower for more massive stars.  Moreover, if
mass segregation has a dynamical origin, we expect to see energy equipartition
between stars of different masses 
(Spitzer \cite{1969ApL....158....139S}).

We can test these hypotheses by splitting the observed  non-binary stars 
into two equal groups of 24 objects according to their 
masses as indicated by their spectral types (Table~\ref{final}).
The result is presented in Table~\ref{masseg}.

\begin{table}[h!]
\caption{Mass segregation}
\label{masseg}
\begin{center}
\begin{tabular}{lrr}
\hline\hline
Mass range &  Mean mass & Velocity dispersion \\ \hline
$>23.5$\msun &  49.6\msun   & 27.8\kms \\
$<23.5$\msun &  19.4\msun   & 36.7\kms \\
\hline
\end{tabular}
\end{center}
\end{table}

The Fischer F-test on both distributions gives a value of F=1.6 corresponding to
a probability of 27\% that both samples are drawn from the same parent distribution.
Thus, there is tentative, but not conclusive, evidence that the massive stars have
a lower dispersion. The ratio of mean energy ($M^2$) between the two mass bins is
$\sim 1.5\pm0.1$, significantly different from the equipartition ratio, $r=1$.
So if the mass segregation is indeed present, it is most likely not due to
two-body relaxation. We remark, however, that our radial velocity data samples
very sparsely the inner 10~pc of the cluster, where we concentrated our photometric
study, and which contains most of the cluster mass.

\section{Conclusions}
\label{sec:remarks}

In spite of the relative low spectral resolution, our data already provide
important new results about the dynamical state of the 30~Dor cluster. First,
the velocity dispersion is much too large to be due random motions of the stars
in the gravitational potential of the cluster. Instead, the observed dispersion
seems to be entirely dominated by binary orbital motions. Thus, the first important
results is that higher spectral resolution alone is not sufficient to probe the
dynamics of the cluster; it is also crucial to have observations with good time
resolution in order to find (and exclude) binaries.  Second, there is no strong
evidence for dynamical mass segregation in the sense of massive stars moving
with lower random velocities.  If present, the effect is masked by binaries, so
again, it is crucial to obtain data for several epochs. Finally, the virial dynamical
mass of the cluster is comparable within factors of a few with the photometric mass.
Therefore, using a reasonable number of non-binary stars it should be possible to
place useful constrains on the IMF slope below 1\msun.

The strong conclusion of this investigation, therefore, is that it would be very
worthwhile to obtain time resolved spectroscopy of a sample of 100-200 stars in
the cluster. The FLAMES integral field spectrograph on the VLT appears ideally suited
for such study.

\end{document}